\documentclass[a4paper,12pt]{article}
\usepackage[utf8]{inputenc}
\usepackage{amssymb}
\usepackage{amsmath}
\numberwithin{equation}{section}
\usepackage{cite}
\usepackage{relsize}

\usepackage{float}
\usepackage[dvipdfmx,setpagesize=false]{hyperref}

\title{Comments on the Nekrasov-type formula \\for E-string theory}
\author{{\large Takenori Ishii}\footnote{gr0054vr@ed.ritsumei.ac.jp}\\
\\
{\small $Department\ of\ Physical\ Sciences,\ Ritsumeikan\ University$}\\
{\small $Shiga\ 525$-$8577,\ Japan$}\\}
\date{}
\begin{document}
\maketitle
\begin{abstract}
We generalise our previous result\cite{13} to cases with three and four Wilson lines.
We show that the Nekrasov-type formula for E-string theory is valid for those cases and the Seiberg-Witten curves are reproduced from the formula.
In addition, we show the dependence of the Seiberg-Witten curve on the Wilson lines.
\end{abstract}

\section{Introduction and Summary}

E-string theory may be the simplest interacting, non-gravitational theory in six dimensions\cite{1,2,3,4,5,6,7,8,9,10}.
It has the (1,0) supersymmetry and the $E_8$ global symmetry.
It contains only a tensor multiplet.
The theory does not have a Lagrangian description but does have rich structures.
One of those is the Seiberg-Witten description\cite{11,12}.
It is obtained by the toroidal compactification down to four dimensions\cite{4,5,6,7} as
\begin{eqnarray}
y^2 &=& 4 x^3 - \frac{1}{12} E_4 ( \tau ) u^4  x - \frac{1}{216} E_6 ( \tau ) u^6 + 4 u^5 ,  \nonumber \\
\frac{ \partial F_0}{\partial \varphi}  &=& 8 \pi^3 i ( \varphi_D - \tau \varphi ) + \mathrm{const.}   ,
\label{eq:1}
\end{eqnarray}
where $E_{4,6} ( \tau )$ are the Eisenstein series with weights 4 and 6, $\tau$ is a complex structure of the torus to which two dimensions were compactified down, $u$ the Coulomb branch moduli parameter, $F_0$ the prepotential, $\varphi$ the Higgs vacuum expectation value(vev), $\varphi_D$ the dual of the Higgs vev, and const. could be a function in $\tau$ but is a constant with respect to $\varphi$.
The toroidal compactification is performed by $\mathbb{C} / \{  z \sim z + 2\pi \sim z+ 2\pi \tau \}$ and gives an $\mathcal{N} =2 \ U(1)$ gauge theory, which is described by (\ref{eq:1}).

Whilst the Seiberg-Witten description gives the rich structure, it is well-known that the Nekrasov-type partition function\cite{9,10,13,14,15,16,17}(for generically $\mathcal{N} =2 \ U(N)$ gauge theories) also gives such a rich structure.
Let $\mathbf{R} = ( R_1 , \cdots , R_N )$ denote an $N$-tuple of partitions.
Each partition $R_k$ is a non-increasing sequence of non-negative integers
\begin{eqnarray}
R_k = \{ \mu_{k,1}  \geqslant \mu_{k,2} \geqslant \cdots , \mu_{k, \ell ( R_k )} > \mu_{k, \ell ( R_k )+1} = \mu_{k, \ell ( R_k )+2}  = \cdots =0 \} .
\label{eq:2}
\end{eqnarray}
Here the number of non-zero $\mu_{k,i}$ is denoted by $\ell ( R_k )$.
$R_k$ is represented by a Young diagram.
Let $| R_k |$ denote the size of $R_k$, i.e. the number of boxes in the Young diagram of $R_k$
\begin{eqnarray}
| R_k | := \sum_{i=1}^\infty \mu_{k,i} = \sum_{i=1}^{\ell ( R_k )} \mu_{k,i} .
\label{eq:3}
\end{eqnarray}
Similarly, the size of $\mathbf{R}$ is denoted by
\begin{eqnarray}
| \mathbf{R} | := \sum_{k=1}^N  | R_k | .
\label{eq:4}
\end{eqnarray}
Let $R_k^\vee  = \{  \mu_{k,1}^\vee  \geqslant \mu_{k,2}^\vee  \geqslant \cdots \}$ denote the conjugate partition of $R_k$ and introduce the notation
\begin{eqnarray}
h_{k,l} (i,j) := \mu_{k,i} + \mu_{l,j}^\vee -i-j+1 ,
\label{eq:5}
\end{eqnarray}
which represents  the relative hook-length of a box at $(i,j)$ between the Young diagrams of $R_k$ and $R_l$.
In the setup, the Nekrasov-type formula for $\mathcal{N} =2 \ U(N)$ gauge theories with $2N$ fundamental hypermultiplets\cite{9,10,13} is given by
\begin{eqnarray}
Z= \sum_{\mathbf{R}} (- e^{2\pi i \varphi} )^{| \mathbf{R} |} \prod_{k=1}^N \prod_{(i,j) \in R_k} \frac{\prod_{n=1}^{2N} \vartheta_1 (  \frac{1}{2\pi} (  a_k - m_n +(j-i)  \hbar ), \tau )}{\prod_{l=1}^N \vartheta_1 ( \frac{1}{2\pi } (  a_k - a_l + h_{k,l} ( i,j ) \hbar ), \tau )^2} ,
\label{eq:6}
\end{eqnarray}
where $\vartheta_1 ( z, \tau )$ is the Jacobi theta function and $m_n$ are the masses of the fundamental hypermultiplets. 
It is worthwhile to mention that for E-string theory $a_k$ are not diagonal components of the Coulomb branch moduli parameter $u$ but just fixed constants as (\ref{eq:8}) or (\ref{eq:9})(for more detail, see\cite{9,10,13}).
For consistency, we require a condition
\begin{eqnarray}
2 \sum_{k=1}^N a_k - \sum_{n=1}^{2N}  m_n =0.
\label{eq:7}
\end{eqnarray}
The condition for the E-string theory with the general Wilson lines is obtained by setting
\begin{eqnarray}
N=4 , \ \  \ a_k = \omega_{k-1} \ \  (k=1,2,3,4) , \ \ \ {\vec{m}}_n = - {\vec{m}}_{n+4} \ \ (n=1,2,3,4).
\label{eq:8}
\end{eqnarray}
Here for $n=1,2,3,4$, ${\vec{m}}_n = - {\vec{m}}_{n+4}$ means $( m_1 , \cdots , m_4 , - m_1 , \cdots , - m_4  )$.
In the case of the E-string theory with the $E_8$ global symmetry, it is obtained by setting all the masses to zeroes, or more simply setting\cite{10}
\begin{eqnarray}
N=3, \ \ \ a_k = \omega_k \ \ (k=1,2,3) , \ \ \ {\vec{m}}_n =- {\vec{m}}_{n+3} =  \vec{0}  \ \ \ ( n=1,2,3 ) ,
\label{eq:9}
\end{eqnarray}
where $\omega_k \ (k=0,1,2,3)$ are half periods of the torus
\begin{eqnarray}
\omega_0 = 0, \ \ \  \omega_1 = \pi , \ \  \ \omega_2 = - \pi - \pi \tau , \ \ \ \omega_3 = \pi \tau  ,
\label{eq:10}
\end{eqnarray}
and ${\vec{m}}_n =- {\vec{m}}_{n+3} = \vec{0}$ means $( m_1 , m_2 , m_3 , - m_1 , - m_2 ,- m_3   )=(0, \cdots ,0)$.
Henceforth, we promise that we always denote only the plus part in the rightmost hand side of the mass relation.
But the reader should remember that the mass relation always includes the minus part.
We proved in the previous work\cite{13}, following the idea of Nekrasov and Okounkov\cite{15}, that the Seiberg-Witten description (\ref{eq:1}) is reproduced from the Nekrasov-type partition function (\ref{eq:6}).
By the toroidal compactification, we have eight Wilson lines at the very most.
When we take $d$ Wilson lines to be non-zeroes, the $E_8$ group breaks to\footnote{The whole broken groups are $E_{8-d} \times A_{d}$.} $E_{8-d} \times U(1 )^{d}$ and these Wilson line parameters are interpreted as the masses of the fundamental hypermultiplets\cite{4,5}.
We also showed that the Seiberg-Witten curves in the cases with one and two specific Wilson lines are reproduced from the Nekrasov-type partition function\cite{13}.    
In this paper, we show that the Seiberg-Witten curves in the general cases with three and four Wilson lines can be reproduced from the Nekrasov-type partition function.
Note that the genuine generality is snatched by our setup (\ref{eq:8}).
At present, it is hard to analyse a case where the eight Wilson lines are not restricted at all.
Under the restriction (\ref{eq:8}) or (\ref{eq:9}), however, we show that the Seiberg-Witten curve for the E-string theory with any feasible, global symmetry is reproduced from the Nekrasov-type partition function.

We make an important comment here.
As we will show in this paper, the Nekrasov-type partition function can reproduce the Seiberg-Witten curve, in other words the prepotential, even in the E-string theory with four Wilson lines.
It cannot, however, give the higher order terms in $\hbar$ at present.
This is due to the modular anomaly equation of the topological string partition function for the local $\frac{1}{2} K3$\cite{9,21}.
The prepotential of the E-string theory with the $E_8$ global symmetry is interpreted as the genus zero topological string amplitude because they give the same modular anomaly equation.
However, their higher order terms do not.
This is why we only discuss the Seiberg-Witten curve.

This paper is organised as follows.
In the next section, we briefly review our previous work\cite{13}, i.e. a flow of the proof for the case with no Wilson lines.
In section three, we firstly show that the Seiberg-Witten curve in the case with three Wilson lines is reproduced from the Nekrasov-type partition function.
In that case, the Seiberg-Witten curve is compared with the one obtained in \cite{19}.
By the comparison, it is seen that not only our result is correct, namely they coincide, but also our result shows the dependence on the Wilson lines explicitly.
In the subsection \ref{subsec:3.2}, we show that the Seiberg-Witten curve in the case with four Wilson lines is reproduced from the Nekrasov-type partition function.
This case is the most general setup in our analysis.
The Seiberg-Witten curve in this case is not explicitly given but we show that a quartic or cubic curve obtained from the Nekrasov-type partition function can lead us to it by following the discussions of the papers \cite{5,19}.
In section four, we make some comments about our analysis and the future works.
We give the conventions of the modular functions used in this paper in Appendix \ref{sec:a}.

\section{Review: the Proof in the Case with no Wilson Lines}

In this section, we briefly review our previous work\cite{13}.
Our purpose is to obtain the Seiberg-Witten description, starting from the Nekrasov-type partition function (\ref{eq:6}).
We here proceed the general story but later focus on reproducing the Seiberg-Witten curve.

Following \cite{14,15}, in the thermodynamic limit $\hbar \to 0$, the Nekrasov-type partition function (\ref{eq:6}) should reproduce the Seiberg-Witten description.
Though the Nekrasov-type partition function is controlled by a set of Young diagrams $\mathbf{R} = \{  R_a , R_2 , \cdots , R_N  \}$, we expect that in the thermodynamic limit it is dominated by some specific Young diagram.
Such an expectation is obtained by rewriting the Nekrasov-type partition function
\begin{eqnarray}
Z \simeq \int D f^{\prime \prime } d^N \lambda \ \exp \Bigg[ \frac{1}{2 \hbar^2} {\mathcal{F}}_0 + \mathcal{O} ( \hbar^0 ) \Bigg] ,
\label{eq:2.1}
\end{eqnarray}
i.e. it is the semi-classical approximation.
Here, $f^{\prime \prime}$ denotes some function which consists of some delta functions.
This function determines the shape of the Young diagram.
$\lambda$ denotes a Lagrange multiplier.
By this, we move on to the next question: the saddle point approximation with respect to $f^{\prime \prime}$.
${\mathcal{F}}_0$ is a functional of the following form
\begin{eqnarray}
{\mathcal{F}}_0 [ f^{\prime \prime} , \lambda_k ] &=& - \frac{1}{2} \int_{\mathcal{C}} dz dw f^{\prime \prime} (z) f^{\prime \prime } (w) \gamma_0 (z-w) + \sum_{n=1}^{2N} \int_{\mathcal{C}} dz f^{\prime \prime} (z) \gamma_0 (z- m_n )  \nonumber \\
&+& 4 \pi i \tilde{\varphi} \Big(  \frac{1}{4} \int_{\mathcal{C}} dz z^2 f^{\prime \prime} (z) - \sum_{k=1}^N \frac{a_k^2}{2}  \Big)  \nonumber \\
&+& 2 \sum_{k=1}^N \lambda_k  \Big(  \frac{1}{2} \int_{{\mathcal{C}}_k} dz z f^{\prime \prime }  (z) - a_k   \Big) ,
\label{eq:2.2}
\end{eqnarray}
where the integrals in the first line are the Cauchy's principal value ones, $\gamma_0 (z)$ is a function which satisfies the difference equation\cite{15}
\begin{eqnarray}
\gamma (z + \hbar ;\hbar ) + \gamma ( z- \hbar ; \hbar ) -2 \gamma (z; \hbar ) = \ln \vartheta_1 \Big( \frac{z}{2\pi}  \Big)
\label{eq:2.3}
\end{eqnarray}
and has the expansion
\begin{eqnarray}
\gamma (z; \hbar ) = \sum_{g=0}^{\infty} \hbar^{2g-2} \gamma_g (z).
\label{eq:2.4}
\end{eqnarray}
However, the explicit form of $\gamma (z; \hbar )$ is not important here and we need just the fact
\begin{eqnarray}
\gamma_0^{\prime \prime} (z) = \ln \vartheta_1 \Big(  \frac{z}{2\pi}  \Big) .
\label{eq:2.5}
\end{eqnarray}
${\mathcal{C}}_k$ denote the local supports around $z= a_k$ and $\mathcal{C}$ denotes their union $\mathcal{C} = \bigcup_{k=1}^N {\mathcal{C}}_k $.
$\tilde{\varphi}$ is defined as
\begin{eqnarray}
\tilde{\varphi} := \left\{
 \begin{array}{l}
\varphi \ \ \ \ \  \mathrm{if}\ N \ \mathrm{is\ odd}  ,\\
\varphi + \frac{1}{2} \  \ \ \mathrm{if} \ N \ \mathrm{is  \ even} .
 \end{array}
 \right. 
\label{eq:2.6}
\end{eqnarray}
From this functional, we obtain the saddle point equation
\begin{eqnarray}
\int_{\mathcal{C}} dw f^{\prime \prime} (w) \gamma_0 (z-w) - \sum_{n=1}^{2N} \gamma_0 ( z- m_n  ) - \pi i \tilde{\varphi} z^2 - \lambda_k z =0 , \ \ \ z \in {\mathcal{C}}_k .
\label{eq:2.7}
\end{eqnarray}
By an analogy of the matrix model, the dominant function $f^{\prime \prime}$ is recovered by a resolvent $\omega (z)$ defined below.
In particular, in the present case, the resolvent is given by an elliptic function $H(z)$ as
\begin{eqnarray}
\omega (z) = \frac{2 \partial_z \sqrt{H(z)}}{\sqrt{H(z) -1}}  . \label{eq:2.8}
\end{eqnarray}
$H(z)$ is given by the Jacobi theta functions\cite{16}
\begin{eqnarray}
H(z) = \kappa \frac{( \prod_{k=1}^N \vartheta_1 ( \frac{z- \zeta_k}{2\pi}  )  )^2}{\prod_{n=1}^{2N} \vartheta_1 ( \frac{z- m_n}{2\pi}  )} , \label{eq:2.9}
\end{eqnarray}
where $\kappa$ and $\zeta_k$ are some constants.

We now arrive at a place where we can reproduce the Seiberg-Witten description.
The integral over the alpha-cycle is given by
\begin{eqnarray}
\frac{\partial \varphi}{\partial u} = \frac{i}{4\pi^2 u}  \oint_\alpha \frac{dz}{\sqrt{1- H(z )^{-1}}} .
\label{eq:2.10}
\end{eqnarray}
The integral over the beta-cycle is discussed in the similar way.
Hence we need to determine the elliptic function $H(z)$ to reproduce the Seiberg-Witten curve (description). 

For the E-string theory with the $E_8$ global symmetry, we set
\begin{eqnarray}
&& N=4 , \ \ \ \ \zeta_k = \omega_k , \ \ \ {\vec{m}}_n =- {\vec{m}}_{n+4} =  \vec{0} , \nonumber \\
\mathrm{or}\ && N=3 , \ \ \ \  \zeta_k = \omega_k , \ \ \  {\vec{m}}_n = - {\vec{m}}_{n+3} = \vec{0}  .
\label{eq:2.11}
\end{eqnarray}
In this case, the elliptic function $H(z)$ becomes
\begin{eqnarray}
H(z) = - \frac{1}{4} u \wp^\prime (z )^2 ,
\label{eq:2.12}
\end{eqnarray}
where $\wp (z)$ denotes the Weierstrass' elliptic function and we have defined\footnote{We can identify $u$ with the Coulomb branch moduli parameter since each size of the three cuts is inversely proportional to $|u|$.}
\begin{eqnarray}
u:= \frac{4\kappa}{q^{1/2} \eta^{12}} , \ \ \ q = e^{2\pi i \tau} .
\label{eq:2.13}
\end{eqnarray}
Then (\ref{eq:2.10}) becomes
\begin{eqnarray}
\frac{\partial \varphi}{\partial u} = \frac{i}{4\pi^2 u} \oint_\alpha \frac{\wp^\prime (z) dz}{\sqrt{\wp^\prime (z )^2 + 4 u^{-1}}} .
\label{eq:2.14}
\end{eqnarray}
Following the Seiberg-Witten description, the Seiberg-Witten curve is given as the Riemann surface of the integrand in the period integral which gives the derivative of the Higgs vev with respect to the Coulomb branch moduli parameter.
Hence we expect that we can identify the Riemann surface of the integrand of (\ref{eq:2.14}) with the Seiberg-Witten curve for E-string theory.
However, this Riemann surface has genus four since the functions are defined on a sheet with three cuts and we have two copies of it.
The Seiberg-Witten curve is, however, an elliptic curve with genus one.
Therefore doing a change of variables
\begin{eqnarray}
\wp (z) = u^{-2} x ,
\label{eq:2.15}
\end{eqnarray}
(\ref{eq:2.14}) becomes
\begin{eqnarray}
\frac{\partial \varphi}{\partial u} = \frac{i}{4 \pi^2}  \oint_{\tilde{\alpha}} \frac{dx}{y} ,
\label{eq:2.16}
\end{eqnarray}
where
\begin{eqnarray}
y^2 = 4 x^3 - \frac{1}{12} E_4 ( \tau ) u^4 x - \frac{1}{216} E_6 ( \tau ) u^6 + 4 u^5
\label{eq:2.17}
\end{eqnarray}
and $\tilde{\alpha}$ is the image of $\alpha$ by the change of variables.
This is exactly the Seiberg-Witten curve for the E-string theory with the $E_8$ global symmetry.

\section{The Generalisation to the Cases with General Wilson Lines}

In this section, we show that the Nekrasov-type partition function (\ref{eq:6}) can reproduces the Seiberg-Witten curves in the cases with three and four Wilson lines.

\subsection{The Case with Three Wilson Lines}

In the case with three Wilson lines, we choose 
\begin{eqnarray}
N=3 , \ \ \ \zeta_k = \omega_k , \ \ \  {\vec{m}}_n = - {\vec{m}}_{n+3} = ( 2\pi m_1 , 2 \pi m_2 , 2 \pi m_3   ).
\label{eq:3.1}
\end{eqnarray}
Then the elliptic function is given by
\begin{eqnarray}
H(z) &=& \kappa \frac{(  \prod_{k=1}^3 \vartheta_1 (  \frac{z- \zeta_k}{2\pi} )  )^2}{\prod_{n=1}^6 \vartheta_1 (  \frac{z- 2\pi m_n}{2\pi} )}  \nonumber \\
&=& \kappa \frac{\vartheta_1 ( \frac{z - \pi}{2\pi}  )^2 \vartheta_1 ( \frac{z + \pi + \pi \tau}{2\pi}  )^2 \vartheta_1 ( \frac{z- \pi \tau}{2\pi}  )^2}{ \vartheta_1 ( \frac{z- 2 \pi m_1}{2\pi}  )\vartheta_1 ( \frac{z-2\pi m_2}{2\pi}  ) \vartheta_1 ( \frac{z -2\pi m_3}{2\pi}  ) \vartheta_1 ( \frac{z+ 2\pi m_1}{2\pi} ) \vartheta_1 ( \frac{z+ 2\pi m_2}{2\pi} ) \vartheta_1 ( \frac{z+ 2\pi m_3}{2\pi} )}  . \nonumber \\  \label{eq:3.2}
\end{eqnarray}
Here, using the identities of Appendix (\ref{eq:a1})-(\ref{eq:a4}) and (\ref{eq:a9}) for the numerator, and using the identity
\begin{eqnarray}
\vartheta_1 \Big( \frac{z+w}{2\pi}  \Big) \vartheta_1 \Big(  \frac{z-w}{2\pi} \Big) = - \eta^{-6} \vartheta_1 \Big( \frac{z}{2\pi} {\Big)}^2 \vartheta_1 \Big( \frac{w}{2\pi} {\Big)}^2 ( \wp (z) - \wp (w)  )  \label{eq:3.3}
\end{eqnarray}
for the denominator, we can express the functions as the Weierstrass $\wp$-functions and obtain
\begin{eqnarray}
H(z) &=&   \frac{\kappa \eta^6 \wp^\prime (z )^2}{ q^{1/2}  \vartheta_1 ( m_1 )^2 \vartheta_1 ( m_2 )^2 \vartheta_1  ( m_3 )^2  ( \wp (z) - \wp ( 2\pi m_1 ) )( \wp (z) - \wp ( 2\pi m_2 ) )( \wp (z) - \wp ( 2\pi m_3 ) )} \nonumber \\
&=&  \frac{u \eta^{18} \wp^\prime (z )^2}{4  \vartheta_1 ( m_1 )^2 \vartheta_1 ( m_2 )^2 \vartheta_1  ( m_3 )^2  ( \wp (z) - \wp ( 2\pi m_1 ) )( \wp (z) - \wp ( 2\pi m_2 ) )( \wp (z) - \wp ( 2\pi m_3 ) )} , \nonumber \\
\end{eqnarray}
where we have used the moduli parameter (\ref{eq:2.13}).
From the discussion in the last section, the integrand of (\ref{eq:2.10}) for the present case becomes
\begin{eqnarray}
\frac{dz}{\sqrt{1- H(z )^{-1}}} = \frac{\sqrt{H(z)} dz}{\sqrt{H(z) -1}} = \frac{\wp^\prime (z )dz}{\sqrt{\wp^\prime (z )^2 - \alpha (m) ( \wp - \wp_1 )( \wp - \wp_2 )( \wp - \wp_3 ) }} . \nonumber \\
\end{eqnarray}
By a change of variables $\wp (z) = x$, we identify this integrand written in terms of the variable $x$ with the Seiberg-Witten curve, i.e. we put
\begin{eqnarray}
y_0^2 &=&  \wp^\prime (z )^2 - \alpha (m) ( \wp (z) - \wp_1  )(  \wp (z) - \wp_2 )( \wp (z) - \wp_3 ) \nonumber \\
&=& ( 4- \alpha ) \wp^3 + \alpha \sigma_1 \wp^2 - ( E_4 + \alpha \sigma_2  ) \wp - ( E_6  - \alpha \sigma_3 )  , \nonumber \\
\label{eq:3.4}
\end{eqnarray}
where
\begin{eqnarray}
\alpha (m) &:=& \frac{4  }{u \eta^{18}} \vartheta_1 ( m_1 )^2 \vartheta_1 ( m_2 )^2 \vartheta_1 ( m_3 )^2 , \nonumber \\
\wp_i &:=& \wp ( 2\pi m_i ) , \nonumber \\
\sigma_1 &:=& \wp_1 + \wp_2 + \wp_3 , \ \ \sigma_2 := \wp_1 \wp_2 + \wp_2 \wp_3 + \wp_1 \wp_3 ,\ \  \sigma_3 :=  \wp_1 \wp_2 \wp_3 ,  \nonumber \\   
\label{eq:3.5}
\end{eqnarray}
and change the variable $z$ into $x$ below.
Here, we have dropped some numerical coefficients, i.e. we have defined $E^\prime_4 := E_4 /12$ and $E_6^\prime := E_6 /216$ and then have dropped the prime.
$u$ is defined as (\ref{eq:2.13}).
Note, however, that this curve is of genus four.
Taking into account that a complex curve we consider is in $\mathbb{C} {\mathbb{P}}^2$, the appropriate change of variables is given by\footnote{We have to pay attention to that the theta functions within $\alpha^{-2}$ and $\alpha^{-3}$ are cancelled, as we will see later.}
\begin{eqnarray}
x:= \alpha^{-2} (4-\alpha ) \wp , \ \ \ y := 2( 4-\alpha ) \alpha^{-3} y_0  . \label{eq:3.6}
\end{eqnarray}
So we can finally obtain the Seiberg-Witten curve
\begin{eqnarray}
y^2 = 4 x^3 - f x -g ,
\label{eq:3.7}
\end{eqnarray}
where
\begin{eqnarray}
f &=&  16 E_4 {\tilde{u}}^4  + ( 16 \sigma_2 - 4 E_4 )  {\tilde{u}}^3   + \Big( \frac{4 \sigma_1^2 }{3} -4 \sigma_2   \Big) {\tilde{u}}^2  , \nonumber \\
g &=& 64 E_6 {\tilde{u}}^6 - \Big( \frac{16 E_4 \sigma_1}{3} +  32 E_6   +  64 \sigma_3     \Big) {\tilde{u}}^5 \nonumber \\
&+& \Big( 4 E_6 + \frac{4 E_4 \sigma_1}{3} - \frac{16 \sigma_1 \sigma_2}{3} +32  \sigma_3 \Big)    {\tilde{u}}^4  
- \Big( \frac{8 \sigma_1^3}{27} - \frac{4 \sigma_1 \sigma_2}{3} + 4 \sigma_3   \Big) {\tilde{u}}^3   , \nonumber \\
\tilde{u} &:=& \frac{u \eta^{18} }{4 \vartheta_1 ( m_1 )^2 \vartheta_1 ( m_2 ) \vartheta_1 ( m_3 )^2} . 
\label{eq:3.8}
\end{eqnarray}
As a check of the consistency, we can compare this with the result obtained in \cite{19}\footnote{Then we have to care about the difference of the notations
\begin{eqnarray}
16 E_4^{\mathrm{ours}} = f_0^{\mathrm{Mohri's}} , \ \ 64 E_6^{\mathrm{ours}} = g_0^{\mathrm{Mohri's}} , \ \ 4 \wp_i^{\mathrm{ours}} = \wp_i^{\mathrm{Mohri's}} ,
\label{eq:3.9}
\end{eqnarray}
and the difference of the number of terms.
We have two more terms $-32 E_6 {\tilde{u}}^5$ and $32 \sigma_3 {\tilde{u}}^4$, and a different coefficient of $-64 \sigma_3 {\tilde{u}}^5$ compared with the Mohri's result (9.18) in \cite{19}.
But we checked that those two terms and the coefficient of (9.18) have been stolen.} and see that our result is in agreement with it.
Therefore it has been shown that in the case with three Wilson lines the Nekrasov-type partition function (\ref{eq:6}) reproduces the Seiberg-Witten curve\footnote{Actually, that the Nekrasov-type partition function reproduces the Seiberg-Witten curve in this case is already seen in the step of (\ref{eq:3.4}) but we have pursued the concrete form of the Seiberg-Witten curve.}.

It is worthwhile to mention that we have here  absorbed the redundant coefficients into the variables as (\ref{eq:3.6}) and the modulus as (\ref{eq:3.8}) to make the form of our result fit to that of \cite{19}.
However, the curve (\ref{eq:3.8}) is obviously divergent when we set one of the Wilson lines to zero.
The expression (\ref{eq:3.8}) lacks more dependence on the Wilson lines.
This originates from the redefinitions of the variables (\ref{eq:3.6}) and the modulus (\ref{eq:3.8}).

To make the dependence on the Wilson lines more explicit, we rescale the Seiberg-Witten curve (\ref{eq:3.7}) by multiplying it by $( 4 \vartheta_1 ( m_1 )^2 \vartheta_1 ( m_2 )^2 \vartheta_1 ( m_3 )^2 / \eta^{18}  )^6$.
Then our result shows the dependence on the Wilson lines explicitly, comparing with the result obtained in \cite{19} based on a prescription of the geometric engineering.
More explicitly, our result is given by
\begin{eqnarray}
{\tilde{y}}^2 &=& 4 {\tilde{x}}^3 - \tilde{f} \tilde{x} - \tilde{g}  , \nonumber \\
\tilde{f} &=&  16 E_4   u^4  + ( 16 \sigma_2 - 4 E_4 ) \alpha^\prime (m )  u^3   + \Big( \frac{4 \sigma_1^2 }{3} -4 \sigma_2   \Big) \alpha^\prime (m )^2  u^2  , \nonumber \\
\tilde{g} &=& 64 E_6 u^6 - \Big( \frac{16 E_4 \sigma_1}{3} +  32 E_6   +  64 \sigma_3     \Big) \alpha^\prime (m ) u^5 \nonumber \\
&+& \Big( 4 E_6 + \frac{4 E_4 \sigma_1}{3} - \frac{16 \sigma_1 \sigma_2}{3} +32  \sigma_3 \Big) \alpha^\prime (m )^2    u^4  
- \Big( \frac{8 \sigma_1^3}{27} - \frac{4 \sigma_1 \sigma_2}{3} + 4 \sigma_3   \Big) \alpha^\prime (m )^3  u^3  , \nonumber \\  \label{eq:3.10}
\end{eqnarray}
where $\alpha^\prime (m) := 4 \eta^{-18} \vartheta_1 ( m_1 )^2 \vartheta_1 ( m_2 )^2 \vartheta_1 ( m_3 )^2$, i.e. $\alpha (m) = \alpha^\prime (m) /u$, and $\tilde{y} := \alpha^\prime (m )^3 y , \ \tilde{x} := \alpha^\prime ( m )^2 x$.
When we set ${\vec{m}}_n = (0,0,0)$, (\ref{eq:3.10}) becomes
\begin{eqnarray}
{\tilde{y}}^2 = 4 {\tilde{x}}^3 - 16 E_4 u^4 \tilde{x} - 64 E_6 u^6 + 256 u^5 .
\end{eqnarray}
Hence we arrive at the Seiberg-Witten curve for the E-string theory with the $E_8$ symmetry\footnote{If the reader wants to make it coincide with (\ref{eq:2.17}) up to the numerical factors, the redefinition ${\tilde{x}}_{new} := {\tilde{x}}_{old} /4$ is required.} (\ref{eq:2.17}).

\subsection{The Case with Four Wilson Lines} \label{subsec:3.2}

In this subsection, we show that the Nekrasov-type partition function in the case with four Wilson lines can reproduces the Seiberg-Witten curve.
This is the most general case in our setup.
In this case, we choose
\begin{eqnarray}
N=4 , \ \ \zeta_k = ( 0, \omega_1 , \omega_2 , \omega_3   ) , \ \ {\vec{m}}_n =- {\vec{m}}_{n+4} = (  2 \pi m_1 , 2 \pi m_2 , 2 \pi m_3 , 2 \pi m_4  ). \nonumber \\
\label{eq:3.11}
\end{eqnarray}
Then the elliptic function $H(z)$ is given by
\begin{eqnarray}
H(z) &=& \kappa \frac{( \prod_{k=1}^4 \vartheta_1 ( \frac{z- \zeta_k}{2\pi}  )  )^2}{\prod_{n=1}^8 \vartheta_1 ( \frac{z- 2\pi m_n}{2\pi}  )} \nonumber \\
&=& \kappa \frac{\vartheta_1 ( \frac{z}{2\pi}  )^2 \vartheta_1 ( \frac{z - \pi}{2\pi}  )^2 \vartheta_1 ( \frac{z + \pi + \pi \tau}{2\pi}  )^2 \vartheta_1 ( \frac{z - \pi \tau}{2\pi}  )^2}{ \vartheta_1 (  \frac{z- 2\pi m_1}{2\pi} ) \cdots \vartheta_1 ( \frac{z - 2\pi m_4}{2\pi}  ) \vartheta_1 ( \frac{z+2\pi m_1}{2\pi}  ) \cdots \vartheta_1 ( \frac{z+2\pi m_4}{2\pi}  )} .
\label{eq:3.12}
\end{eqnarray}
By performing repeatedly the procedure in the last subsection, we obtain
\begin{eqnarray}
y_0^2 &=&  \wp^\prime (z )^2 + \alpha (m) ( \wp (z) - \wp_1   ) ( \wp (z) - \wp_2   )( \wp (z) - \wp_3 ) ( \wp (z ) - \wp_4 )  \nonumber \\
&=& 4 \wp^3 - E_4 \wp - E_6 + \alpha (  \wp^4 - \sigma_1 \wp^3 + \sigma_2 \wp^2 - \sigma_3 \wp + \sigma_4  ) , \nonumber \\
\label{eq:3.13}
\end{eqnarray}
where
\begin{eqnarray}
\alpha (m) &:=& \frac{4 }{u \eta^{24}  } \vartheta_1 ( m_1 )^2 \vartheta_1 ( m_2 )^2 \vartheta_1 ( m_3 )^2  \vartheta_1 ( m_4 )^2  , \nonumber \\
\sigma_1 &:=& \wp_1 + \wp_2 + \wp_3 + \wp_4  , \nonumber \\
\sigma_2 &:=& \wp_1 \wp_2 + \wp_2 \wp_3 + \wp_3 \wp_4 + \wp_1 \wp_3 + \wp_1 \wp_4 + \wp_2 \wp_4  , \nonumber \\
\sigma_3 &:=& \wp_1 \wp_2 \wp_3 + \wp_1 \wp_2 \wp_4 + \wp_1 \wp_3 \wp_4 + \wp_2 \wp_3 \wp_4 , \nonumber \\
\sigma_4 &:=& \wp_1 \wp_2 \wp_3 \wp_4 .
\label{eq:3.14}
\end{eqnarray}
Though the curve (\ref{eq:3.13}) is superficially a quartic curve, by restoring the homogeneous coordinates as
\begin{eqnarray}
x_1 = \wp , \ \ x_2 = y_0, \ \ x_0 \neq 1 ,
\label{eq:3.15}
\end{eqnarray}
and recalling the fact that the curve is
\begin{eqnarray}
x_0 x_2^2 = 4  x_1^3 - E_4 x_0^2 x_1 - E_6 x_0^3
\label{eq:3.16}
\end{eqnarray}
at $u= \infty$, we can obtain the cubic curve
\begin{eqnarray}
( a_0 x_0 - \alpha x_1   ) x_2^2 &=& 16 x_1^3 + a_1 x_0 x_1^2 + a_2   x_0^2 x_1 + a_3  x_0^3 ,
\label{eq:3.17}
\end{eqnarray}
where
\begin{eqnarray}
a_0 &:=& 4 + \alpha \sigma_1 , \nonumber \\
a_1 &:=& \alpha E_4 + 4 \alpha \sigma_2 , \nonumber \\
a_2 &:=& -4 E_4 + \alpha E_6 - \alpha E_4 \sigma_1  - 4 \alpha \sigma_3 , \nonumber \\
a_3 &:=& -4 E_6 - \alpha  E_6 \sigma_1 + 4 \alpha  \sigma_4 .
\label{eq:3.18}
\end{eqnarray}

One can confirm that (\ref{eq:3.17}) coincides with that given in \cite{19} under the suitable identifications of parameters, which can be rewritten in the Weierstrass form with the discriminant expected in the E-string theory, according to the general arguments given in \cite{5,19} (especially, by the formula presented in Appendix A of \cite{19}). 
It means, thus, that the Nekrasov-type partition function can reproduce the Seiberg-Witten curve in the case with four Wilson lines.

\section{Remarks}

In this paper, we showed that the Nekrasov-type partition function can also reproduces the Seiberg-Witten curve in more general cases.
In particular, the case with four Wilson lines is the most general in our analysis.
We hope that an analysis which is more general, i.e. not restricted as ${\vec{m}}_n = - {\vec{m}}_{n+4}$, will appear elsewhere.
In addition, we also showed explicitly the dependence of the Seiberg-Witten curve on the Wilson lines in the case of three Wilson lines.
This is an explicit result obtained from the Nekrasov-type partition function, not obtained from the method of the geometric engineering.

Finally, we make a comment about the reproduction of the Seiberg-Witten curve.
The generalised curves should include all results we know, e.g. \cite{10,13}.
We have seen that the curve (\ref{eq:3.7}) is reliable by comparing it with the result obtained in \cite{19}.
We did not explicitly give the most general Seiberg-Witten curve, i.e. the Seiberg-Witten curve in the case with four Wilson lines, but as a consistency check of our result, the most general curve should include the curve (\ref{eq:3.7}).
This is easily shown.
As mentioned in subsection 3.2, we can say that the quartic curve (\ref{eq:3.13}) is correct by comparing it with the quartic curve given in \cite{19}.
When we set one of the Wilson lines to zero, by the term $\lim_{m\to 0} \wp (2\pi m ) \vartheta_1 ( m )^2 = \eta^6$ in the curve (\ref{eq:3.13}), we get the curve (\ref{eq:3.7}).
Note that each case is reproduced by setting the Wilson lines to the appropriate values.

\begin{center}
{\bf Acknowledgements}
\end{center}

The author is deeply indebted to Yuji Sugawara for valuable discussions.
And also the author thanks Kazuhiro Sakai for leading him to the world of E-string theory.

\appendix
\section{Conventions of special functions}  \label{sec:a}

The Jacobi theta functions are defined as
\begin{eqnarray}
\vartheta_1 (z , \tau ) &:=& i \sum_{n\in \mathbb{Z}} (-1 )^n y^{n-1/2} q^{(n-1/2 )^2 /2} , \label{eq:a1} \\
\vartheta_2 (z , \tau ) &:=&  \sum_{n\in \mathbb{Z}}  y^{n-1/2} q^{(n-1/2 )^2 /2}  , \label{eq:a2} \\
\vartheta_3 (z , \tau ) &:=&  \sum_{n\in \mathbb{Z}}  y^{n} q^{n^2 /2}  , \label{eq:a3} \\
\vartheta_4 (z , \tau ) &:=&  \sum_{n\in \mathbb{Z}} (-1 )^n y^{n} q^{n^2 /2}  ,\label{eq:a4}
\end{eqnarray}
where $y= e^{2\pi iz} , \ q = e^{2\pi i\tau}$.
We often use the following abbreviated notation
\begin{eqnarray}
\vartheta_k (z) := \vartheta_k (z, \tau), \ \ \  \vartheta_k := \vartheta_k (0, \tau ) .\label{eq:a5}
\end{eqnarray}
The Dedekind eta function is defined as
\begin{eqnarray}
\eta (\tau ) := q^{1/24} \prod_{n=1}^\infty (1- q^n )  . \label{eq:a6}
\end{eqnarray}
The Eisenstein series are given by
\begin{eqnarray}
E_{2n} ( \tau ) = 1+ \frac{2}{\zeta (1-2n) } \sum_{k=1}^\infty \frac{k^{2n-1} q^k}{1- q^k} .
\end{eqnarray}
We often abbreviate $\eta ( \tau ), \ E_{2n} ( \tau )$ as $\eta , \ E_{2n}$ respectively.

The Weierstrass $\wp$-function is defined as
\begin{eqnarray}
\wp (z) = \wp (z ; 2 \omega_1 , 2 \omega_3 ) := \frac{1}{z^2} + \sum_{(m,n) \in {\mathbb{Z}}^2_{\neq (0,0)}} \Big[   \frac{1}{(z- \Omega_{m,n} )^2} - \frac{1}{\Omega_{m,n}^2}  \Big]  , \label{eq:a7}
\end{eqnarray}
where $\Omega_{m,n} = 2m \omega_1 + 2n \omega_3$ and
\begin{eqnarray}
\omega_1 + \omega_2 + \omega_3 =0 , \ \ \  \frac{\omega_3}{\omega_1} = \tau  .\label{eq:a8}
\end{eqnarray}
In the main text, we use the following identities
\begin{eqnarray}
\wp^\prime (z )^2 &=& 4 \wp (z )^3 - \frac{\pi^4}{12 \omega_1^4} E_4 \wp (z) - \frac{\pi^6}{216 \omega_1^6} E_6  \nonumber \\
&=& 4 ( \wp (z) - e_1 )( \wp (z) - e_2 )( \wp (z) - e_3 )  \nonumber \\
&=& \frac{\pi^6}{\omega_1^6} \eta^{12} \prod_{k=1}^3  \frac{\vartheta_{k+1} (  \frac{z}{2\pi}  )^2}{\vartheta_1 ( \frac{z}{2\pi}  )^2}  , \label{eq:a9}
\end{eqnarray}
where $e_k := \wp ( \omega_k )$.

\end{document}